\documentclass[10pt,psfig,letterpaper,twocolumn]{article}

\usepackage{abstract}
\usepackage{geometry}
\usepackage{graphics}
\usepackage{setspace}

\usepackage{amsmath,amsfonts,amssymb,slashed,cancel}
\usepackage{amsthm}
\usepackage{graphicx}
\usepackage{epsfig}
\usepackage{graphics}
\RequirePackage{graphicx}
\usepackage{hyperref}
\usepackage[small]{caption}
\usepackage{float}
\usepackage{wrapfig}
\date{}
\newcommand{\be}{\begin{equation}}
\newcommand{\ee}{\end{equation}}
\newcommand{\bi}{\begin{itemize}}
\newcommand{\ei}{\end{itemize}}

\title{\textbf{Demystifying the Delayed Choice Experiments}}
\author{Bram Gaasbeek\\ \textit{Institute for Theoretical Physics, K.U.Leuven, Celestijnenlaan 200D, 3001 Leuven, Belgium}}%bramg
\begin{document}

\twocolumn[
\maketitle  
\begin{onecolabstract}
The delayed choice experiments are a collection of experiments where the counterintuitive laws of quantum mechanics are manifested in a very striking way. They most definitely fit in the legendary series of situations where the underlying quantum nature of reality is `thrown in your face' \cite{Coleman}. Although the delayed choice experiments can be very accurately described with the standard framework of quantum optics, a more didactical and intuitive explanation seems not to have been given so far. In this note, we fill that gap. \\
\end{onecolabstract}
]

\thispagestyle{empty}
 
%\newpage
\section{Invitation}

The EPR-paradox \cite{EPR}\cite{Bohm} and subsequent work of Bell \cite{Bell} are without doubt %one of the most important 
one of the cental pillars of the modern understanding of quantum mechanics. Roughly speaking, 
the generally accepted implications are: (1) the predictions of quantum mechanics can not be reproduced by a hidden variable theory, and (2) quantum mechanics is inherently nonlocal. In this introduction, we will focus our attention on the issue of nonlocality. 

Consider a setup where particles $A$ and $B$ are placed at spatially separated experimenters Alice and Bob. The particles are in the state:
\be
\frac{|\uparrow\rangle _A|\downarrow\rangle _B \quad -  \quad |\downarrow\rangle _A|\uparrow\rangle _B}{\sqrt{2}}.
\ee
Now they measure, say Alice does this first. A well-known peculiarity is that they will find opposite outcomes (one finds up, the other finds spin down) even if Bob measures the particle right after Alice did, \textit{before any causal signal can reach him}. The standard one-liner that resolves the paradox is the fact that `correlation is not causality'. (Which is also a crucial fact in the domain of Statistics.) Even though their two measurements are fully correlated (knowing one means knowing the other) they are not causally related \cite{Sakurai}. That is: neither can choose the outcome of their measurement. Otherwise, they could influence the measurement of the other at will, and hence communicate instantly.

\subsubsection*{From forward light cone to collapse}
\begin{figure}
\begin{center}
\includegraphics[width=80mm]{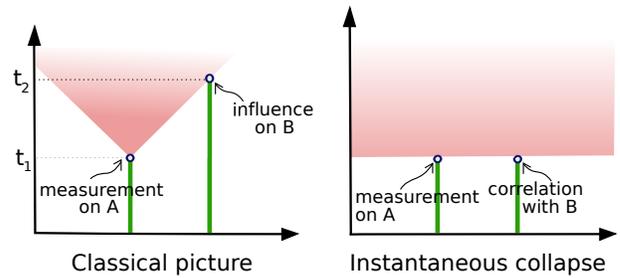}
\caption{Left: %According to classical (relativistic) physics
Naively, one would expect the influence of a measurement to be contained within the forward lightcone emerging from that event. Right: The picture of an instantaneous collapse. As explained in the text, this is not in contradiction with causality. Note: in this work, we will not make any reference to \textit{why} this (apparent) collapse occurs. Not only is this a much harder issue, it is simply not relevant to discussion we will present.
}
\label{fig:EPR}
\end{center}
\end{figure}

The consequence of the above paradox and its resolution is not just philosophical, but very real. It shows that measuring entangled but distant particles results in an \textit{instantaneous} and \textit{overall} collapse, %(instead of a lightcone-like influence) 
yet without violating (relativistic) causality. This is pictorially summarized in Figure \ref{fig:EPR}.

\subsubsection*{The subtlety}

\begin{figure}
\begin{center}
\includegraphics[width=70mm]{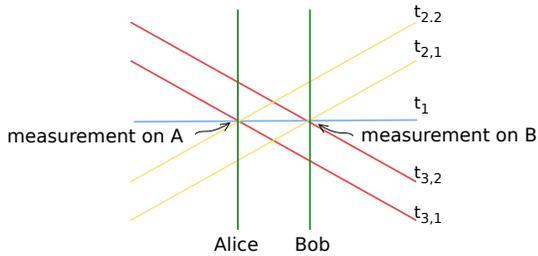}%{figure2_recolor}
\caption{According to special relativity, moving observers have `twisted' reference frames, with tilted time and space axes. Here, lines of constant time are drawn for three different reference frames. Blue lines: rest frame of Alice and Bob. Red and yellow lines: frame of an observer moving to the left or right respectively.  %If Alice and Bob measure simultaneously (in their frame), they get opposite results. So for example: Bob finds up and Alice down. 
%The two boosted observers see a different order of events. 
According to frame 2, Bob measures first. When he does so, particle $A$ collapses too, so this happens on slice $t_{2,1}$. If Alice measures particle $A$ an instant later (on time $t_{2,2}$), she necessarily finds an opposite result. However, according to frame 3, it is Alice who measures first. So according to this observer the reduction occurs along line $t_{3,1}$ and it is Bob who is left with a collapsed particle at a later time $t_{3,2}$.}
\label{fig:EPR_boost}
\end{center}
\end{figure}

The standard resolution of the locality paradox, as sketched above, seems very satisfactory. However, there is a subtlety involved \cite{Peres}. %From the point of view of special relativity, 
It is rather unclear on exactly which moment a particle should experience a co-collapse when its entangled twin is subjected to a measurement. 
%Indeed: relativity tells us that %(unless nature would single out the rest frame of $A$ and $B$ as special) 
%there is no natural `simultaneous moment' for $B$ to collapse together with $A$. 
More concretely: imagine Bob and Alice measure their spins at exactly the same moment according to \textit{their} reference frame. %Alice measures `down'. 
For an observer moving to the left (right) however, it is Alice (Bob) who measures first and thus causes the collapse of the other particle. So the question 
\begin{quote} `When does the co-collapse of an entangled \\ particle occur?' \end{quote}
is solved in a very curious way. Different observers plainly \textit{disagree} on the moment at which the co-collapse occurs (see Figure \ref{fig:EPR_boost}) but they all see an order of events which is in perfect concordance with causality. %the standard resolution of the EPR paradox. 
\newpage
So in the experiment at hand, the (hard) question on \textit{when} the wave function reduction of the entangled particle occurs somehow seems to be irrelevant. However, one might be suspicious. Maybe things do not always work out so nicely. This is precisely where the delayed choice experiments come in. We briefly review these experiments, and show that essentially the above question/problem re-appears, but now in a matured version.

\section{Delayed choice experiments}

Originally, the notion of `delayed choice' arose in a set of thought experiments, devised by Wheeler \cite{Wheeler}. Since then, several variants have found experimental realization \cite{Eraser}\cite{Cont}\cite{Cont2}. For concreteness, we shall restrict our attention to one specific (slightly alternative but more instructive) realization, the \textit{delayed choice quantum eraser}.  
A discussion of another (more literal, but less instructive) variant can be found in Appendix A.

\subsubsection*{The delayed choice quantum eraser experiment}
\begin{figure}
\begin{center}
\includegraphics[width=80mm]{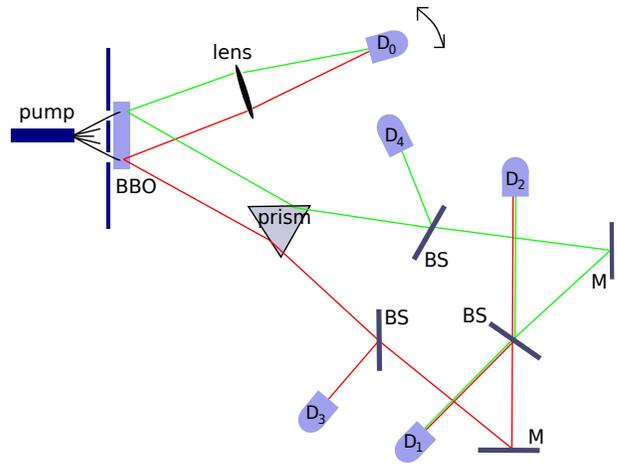}
%quantum eraser - \rangle  HEET fig 4 in tekst maar is fig 5 in bestand. -\rangle  fig 4 is verwijderde wheeler. -\rangle  hernoemd naar fig4noWh.
\caption{The delayed choice quantum eraser experiment. A pump laser beam shines on a double slit. A BBO crystal splits each photon in two entangled photons. The (moveable) detector $D_0$ detects the photon traveling upwards (`signal photon') and can scan over various positions to detect an interference pattern - as if a detection screen were present. The photon going downwards (`idler photon') is received by a prism, and a set of beamsplitters (BS) and mirrors (M). If the idler is detected by $D_3$ or $D_4$, it can only have come through one of the two slits. If it is detected by $D_1$ and $D_2$, it may have travelled via either of the two ways, and does not reveal any which-path information. The arrangement is such that the detection of each signal photon always occurs \textit{before} the detection of the corresponding idler photon. 
}
\label{fig:eraser}
\end{center}
\end{figure}

The delayed choice quantum eraser experiment was first realized in \cite{Eraser}. The details are shown in Figure \ref{fig:eraser}. The outcome is as follows: signal photons for which the corresponding idler photon later reveals which-path information, do not show an interference pattern. Their detection rates are precisely those of collapsed, single slit paths. Signal photons for which the idler does not reveal any path-information, form an untouched interference pattern. So interference at $D_0$ only occurs for events where the idler photon is detected at $D_1$ or $D_2$.
\newpage

\subsubsection*{Interpretation}
In this experiment, the collapse of the wave function (seen in the pattern of the signal photons) is induced by an event (the detection of the idler photon, if it occurs at $D_3$ or $D_4$ at least) which happens at a \textit{later} moment in time. This is a very striking result. How can a measurement cause an entangled particle to collapse \textit{in the past}? It seems like the picture of an instantaneous wave function collapse, as sketched in the first sections, has to be wrong. One might be drastic and postulate a form of `backwards time influence', where a measurement may influence the `behavior' of particles (or here, photons) in the past \cite{BTI}\cite{ShortTime}.
%\footnote{This is the same conclusion one would naively draw from Wheeler's original thought experiment, discussed in the appendices.} 
\textbf{It is precisely this (common) confusion which we wish to clear out in this note.}
%Is there an alternative to this awkward conclusion? Luckily, yes. %It turns out there is one, which will at the same time answer the question we started out with. 

First of all, note that a quantitative computation (using standard quantum optical techniques) reproduces precisely the experimental outcome \cite{Eraser}. This result follows without invoking any backwards causality, and clearly indicates that the experiment can be understood within the standard quantum mechanical framework.

%out introducing new physics, or any sort of a `backwards time interpretation' \cite{BTI}.

But what is going on then? How can we understand this experiment in a more \textit{intuitive} way and what are the conceptual conclusions? % we can draw from this apparent contradiction? 
We will resolve these issues in the next sections.

\newpage
\section{Conditional probabilities}
To come to a resolution, we first obtain some results on conditional measurement probabilities. Say we have two entangled (but spatially separated) particles $A$ and $B$, and we want to measure observables $\mathcal{O}_A$ and $\mathcal{O}_B$. If we have corresponding eigenbases $|i\rangle _A $ and $|j\rangle _B $, we can write a general state as
\be
|\psi\rangle =\sum_{i j} \alpha_{ij} |i\rangle _A |j\rangle _B
\ee
with $\sum_{i j} |\alpha_{ij}|^2=1$. For simplicity, we assume that the above states all have the same energy. Time evolution then involves an irrelevant overall phase factor, which we suppress. When measuring $\mathcal{O}_A$, the chance that the outcome $O_A$ equals the $I$-th eigenvalue $a_I$ is given by
\be
P(O_A=a_I)=\sum_{j} |\alpha_{Ij} |^2.
\label{eq:original prob}
\ee
And similar for $P(O_B=b_J)$. Say we have indeed measured on $B$ and got $O_B=b_J$. The state then collapses onto %, but not performed any measurement on $B$ yet. 
\be
|\psi\rangle \rightarrow |\psi'\rangle =\frac{\sum_i \, \alpha_{iJ}|i\rangle |J\rangle}{\sum_{i} |\alpha_{iJ}|^2}.
\ee
and for this $|\psi'\rangle $ the chance to measure $O_A=a_I$ is
\be
P(O_A=a_I|O_B=b_J)= \frac{|\alpha_{IJ}|^2}{\sum_{i} |\alpha_{iJ}|^2}.
\label{eq:new prob}
\ee
In words: measuring $\mathcal{O}_B$ has resulted in a change in probabilities (\ref{eq:original prob}) $\rightarrow$ (\ref{eq:new prob}).

So the combined chance to first find $O_B=b_J$ and then $O_A=a_I$ is given by 
\be
P(O_A=a_I|O_B=b_J) P(O_B=b_J)
%P(O_B=b_J|O_A=a_I) P(O_A=a_I)
%\label{eq:the other}
\ee
Now what is the chance for this to happen in the other order? That is, what is the chance to \textit{first} measure $O_A=a_I$, and then find $O_B=b_J$? This is given by
\be
P(O_B=b_J|O_A=a_I) P(O_A=a_I)
\ee
And the last two expressions are equal by Bayes' theorem. (Both are of course just equal to $|\alpha_{IJ}|^2$.) 

None of this looks very surprising, but we want to stress that the total probability to find $O_A=a_I$ and $O_B=b_J$ \textbf{does not depend on the place or time} at which the measurements occur. Before we interpret this simple but important result, we generalize it a to performing \textit{several} measurements on the two particles.
So consider observables $\mathcal{O}_{A,1}\cdots\,\mathcal{O}_{A,n}\,\,$ and $\,\,\mathcal{O}_{B,1}\cdots\,\mathcal{O}_{B,m}\,$. 
%The measurements on each particle don't have to be compatible, so the $[\mathcal{O}_{A,i},\mathcal{O}_{A,j}]$ are not necessarily zero, and similar for $B$. 
%However, the Hilbert space is still of product form, so we have $[\mathcal{O}_{A,i},\mathcal{O}_{B,j}]=0$ for all $i$ and $j$. 
We still suppose $A$ and $B$ are each in an energy eigenstate, so that time evolution is trivial. Now consider for each operator \textit{one} specific eigenvalue. We denote these by $\lambda^A_i$ and $\lambda^B_j$. If we start with a joint state $|\psi\rangle $ and measure $\mathcal{O}_{A,1}\cdots\,\mathcal{O}_{A,n}\,\,$ and $\,\,\mathcal{O}_{B,1}\cdots\,\mathcal{O}_{B,m}\,$ (in that order) then the chance to obtain as outcomes the numbers $\{\lambda^A_1\cdots\lambda^A_n,\lambda^B_1\cdots\lambda^B_m\}$ is given by
\be
P(\lambda^A_1\cdots\lambda^A_n,\lambda^B_1\cdots\lambda^B_m) = \langle \psi|P_{\lambda^A_1}\cdots P_{\lambda^A_n}P_{\lambda^B_1}\cdots P_{\lambda^B_m}|\psi\rangle 
\ee
Where $P_{\lambda^A_i}$ and $P_{\lambda^B_j}$ are the projectors on the eigenspaces corresponding to the eigenvalues $\lambda^A_i$ and $\lambda^B_j$. Now the measurements on each particle don't have to be compatible, so the $[P_{\lambda^A_i},P_{\lambda^A_j}]$ are not necessarily zero, and similar for $B$. However, the product structure of the Hilbert space implies $[P_{\lambda^A_i},P_{\lambda^B_j}]=0$ for all $i$ and $j$. So in the right hand side, we can freely move the $P_{\lambda^B}$ through the $P_{\lambda^A}$, as long as we preserve the order of the $P_{\lambda^B}$ and the $P_{\lambda^A}$ amongst themselves. For example, we can drag $P_{\lambda^B_1}$ completely to the left. This implies that
\be
P(\lambda^A_1\cdots\lambda^A_n \lambda^B_1\lambda^B_2\cdots\lambda^B_m) =  P(\lambda^B_1\lambda^A_1\cdots\lambda^A_n \lambda^B_2\cdots\lambda^B_m)
\ee
In words: the probability of a collection of outcomes does not depend on the relative moments of measurements on $A$ and $B$. As long as we respect the order of measurements on each particle separately, the chance stays the same. This generalizes the previous result.% to several measurements.

\subsubsection*{Interpretation}
Although the above expressions are all very simple, the result is, upon second thought, very non-trivial. 
It shows that in general, the relative time ordering of measurements on separated (but possible entangled) 
particles $A$ and $B$ doesn't matter at all. (Of course, for measurements on the same particle, the order 
\textit{does} matter.) %This seems to confirm our first naive guess in the EPR context: namely 
%that one can define the collapse of an entangled pair of particles to occur along any spatial slice -
% which one you choose does not matter anyway. But the conclusion is actually stronger. 
What does this say about the collapse of the total wave function? To see this, it might be elusive to write
the projection operators from the previous section as their action on the product Hilbert space 
$\mathcal{H}_A\otimes \mathcal{H}_B$ of the two particles:
\begin{eqnarray}
P_{\lambda_i^A} \rightarrow P_{\lambda_i^A}& \otimes &1\\
P_{\lambda_i^B} \rightarrow\,\,\,1\,\,\,&\otimes & P_{\lambda_i^B}
\end{eqnarray}
This makes explicit that a measurement  on one particle does \textit{not at all} influence the other one. (I.e. the operator $1$ acts trivially.) 
The only effect a measurement has, is changing probabilities of other measurements into conditional probabilities, as explained just above. 
More important, these conditional probabilities hold regardless of the moment at which you perform the
measurement on the other particle. Whether it occurs later, earlier or at the same time - that doesn't matter at all. 
This forces us to abandon the (popular, but incorrect) view on the 
wave function collapse as an event stretching out along a space-like slice. Even though this view is appealing, it creates a wrong intuition about the physics involved.% and actually it is the source of confusion in the delayed choice experiments.
\newpage
%Below, we will show that the confusion around the delayed choice experiments can be avoided with the point stressed above. First, we use it to
We can now solve the problem we started with in the introduction. If the measurement is nothing but an isolated event in space time, there is no point whatsoever in trying to associate a spatial slice to it. %Indeed, relativity tells that there is no preferred slice of `constant time' passing through one single event. 
So the horizontal and tilted lines in Figure \ref{fig:EPR_boost} actually have no meaning at all! 
Nothing happens \textit{along} these slices - the only place where something physical happens is the place 
of the measurement, and the implications on conditional probabilities hold for other measurements throughout the entire spacetime, \textit{present and past}. %Having cleared out this issue, we can now give a simple view on the delayed choice quantum eraser experiment.

%misschien projectoren nog uitschrijven als product met eenheid.
%The probability for this to happen is the same, regardless the order of the measurements,
 %whether the measurements are spacelike or timelike separated, etcetera. 
%This is, upon second thought, a very non-trivial fact. 

\subsection{Delayed choice quantum eraser, ctd.}

Having answered our initial question, we now turn to the main goal of this note: understanding the delayed choice quantum eraser experiment in a more intuitive way. 
We can write a schematic time evolution, as follows. The initial photon state $|I>$ (or `photon wave function' \cite{Photonwave function}) 
% in 
%Heisenberg notation. 
%We write the initial photon from the pump as follows:
%\be
%| I \rangle 
%\ee
falls through the slit %, the state (or `photon wave function' \cite{Photonwave function}) 
and then consists of an upper and a lower part:
\be
|I\rangle \rightarrow \frac{1}{\sqrt{2}} (|U\rangle + |L\rangle)
\ee
%Actually, the above `transition' - like all the following ones - is nothing but a change of basis, so we could write an equality. 
Each of the two components of the wave function will then be split into an entangled pair of photons: one traveling upward ($\nearrow$) and one downward ($\searrow$)
\be
\rightarrow  \frac{1}{\sqrt{2}} \left( |U \nearrow \rangle |U \searrow \rangle+ |L \nearrow \rangle |L \searrow \rangle \right)
\ee
The idler photon ($|U \searrow \rangle$ and $|L \searrow \rangle$) meets some beamsplitters, 
and is directed to the four detectors. Denoting the components heading to detectors $D_1$ to $D_4$ by their respective numbers, we get the state
\begin{align}
|U \nearrow \rangle & \left(\frac{1}{2} |4\rangle +
 \frac{1}{2\sqrt{2}} |1\rangle + \frac{1}{2\sqrt{2}} | 2 \rangle \right) \nonumber \\
+ & |L \nearrow \rangle 
\left(\frac{1}{2} |3 \rangle + \frac{1}{2\sqrt{2}} |1\rangle + \frac{1}{2\sqrt{2}} | 2 \rangle \right)
\end{align}
The prefactors %follow from the specific experimental configuration, and 
can be checked by looking at Figure \ref{fig:eraser}.
For simplicity, we have not taken into account polarization issues and the sign changes due to reflections. 
We now use what we have learned above. The time ordering of the detections of the idler and signal photon
does not influence at all the probability of a certain outcome. So the experimental outcome (encoded in 
the combined measurement outcomes) is bound to be the same even if we would measure
the idler photon earlier, i.e. \textit{before} the signal photon by shortening the optical path length of the downwards configuration. Then, if the idler is detected at $D_4$ for example, the above state `collapses' onto 
\be
|4\rangle  |U \nearrow \rangle
\ee
which means the signal photon gets only one contribution and hence there can not be any interference. If the idler photon is detected at $D_1$ however, 
the state `collapses' onto 
\be
\frac{1}{\sqrt{2}}|1 \rangle (|U \nearrow \rangle + |L \nearrow \rangle)
\ee 
so there are two contributions to the signal photon and interference occurs. 
Similarly $D_2$ gives interference and $D_3$ does not. This indeed reproduces the experimental outcome correctly, and explains the experiment in a rather concise way.

In the above, we have put \textit{collapse} between quotation marks, 
as the previous section stressed that it is a local (and not global) event which
is correlated with other outcomes regardless the relative time ordering.
In conclusion: nor `causality' nor `influence' nor `collapse' are good words in the context of measurements, 
only `correlation' and `conditional probability' are.

A remark on assumptions. In the previous section, we supposed the measured observables to be conserved. This is necessary to carelessly time-translate the projection operators. The translated observable here is the idler photon measurement. This determines the beam of photon (so its momentum) and is clearly conserved.

\section{Conclusion}

In this note, we discussed the delayed choice experiments. Specifically, we focussed on the role of conditional probabilities for measurements on entangled particles. 

%we have been able to understand the counterintuitive delayed choice phenomenon in a more straightforward manner.
%One of the historical lessons we drew from the EPR thought experiment 
The (well-known) point stated in the introduction was to distinguish correlation from causation. The lesson we draw here is that this very correlation between distant measurements does not feel their relative time ordering: it does not distinguish between future and past. This implies backwards \textit{correlation} but still precludes backwards \textit{causation} or any other tension with relativity, effectively demystifying the delayed choice experiments.

It is important to note that arriving at our conclusions did not require introducing new physics. We only relied on elementary quantum mechanics: not on novel `backwards time' concepts, nor on any particular interpretation: we only used the Born rule `as is'. And it better be so, since a careful quantum optical analysis of any of the delayed choice experiments is in perfect concordance with experimental results - without any auxiliary/new input. Only, these quantum optical analyses are slightly less transparent, and may leave some conceptual issues unclear and confusing. It is precisely this gap that we intended to fill with this note. With the remarks and intuition presented here, there really is no mystery whatsoever in any of the discussed experiments.

\subsubsection*{Acknowledgements}
The author wishes to thank F. Denef, B. Vercnocke and especially F. de Melo for comments on the manuscript.

\newpage

\section*{Appendix A: Wheeler's thoughts}
Here, we present Wheeler's original thought experiment. To study the collapse of the wave function, he devised the following setup \cite{Wheeler}. Imagine a light source (or for the same matter: a particle beam) aimed at a double slit. At a far distance, a detector screen is placed. Obviously, an interference pattern forms on this detector screen. Now place two `telescopes' behind the screen, each tightly aimed at one of the two slits. %Also, allow the detecting screen to be pulled away at will. 
If the screen is pulled away, the telescopes allow to detect through which of the two slits a particle came, %. Indeed, each particle can only be detected in one telescope 
revealing which-slit information and suggesting a point particle behavior. So by putting the screen in, or pulling it away, an experimenter can (at the very last instance) choose to reveal either the wave - or the particle nature of a particle coming through the slits. 
%IRRELEVANT: The setup is also shown in Figure \ref{fig:Wheeler}.
%\footnote{You may be a bit suspicious whether it is possible to aim two telescopic detectors tightly enough to distinguish through which slit the light came. A way to see that this can be done -in principle- is by considering situations of astronomic size. Imagine a light source standing behind a heavily gravitating object. The }
%\subsubsection*{Difference with ordinary two slit experiment}

The essence of Wheeler's variation on the double slit experiment, lies in the words `at the very last instance'. It is a well known fact that trying to detect the path of a particle collapses the wave function and hence destroys the interference pattern. %. The most common interpretation of this phenomenon is 
What happens in Wheeler's experiment is different. As you choose to measure the particle- or wave aspect only at the \textit{very last instant}, the `decision' to act like a wave or a particle is made long \textit{after} it passed the slits. In some sense, a measurement in the present seems to influence an event in the past. Recently, a rather literal version of Wheeler's experiment has been realized \cite{ExpWheeler}, confirming the counterintuitive outcome described above. 
\subsubsection*{Confusing explanation}
One may be tempted to conclude that a measurement can influence the behavior of a particle \textit{in the past}. Measuring the interference pattern (detector screen) enforces the wave character, measuring which-slit information (telescopes) retrojects into a particle behavior. %This influence of a measurement on the past (= the traveling time between the slits and the screen) makes the notion of complementarity even more weird. 
Conceptually, this is a highly confusing explanation, and definitely suggests some form of backwards causality. Luckily, this is not the end of the story.

\subsubsection*{Less confusing explanation}
%It has been stated at several places that the above interpretation is not strictly neccesary. %Obviously, it is not justified to infer from a measurement the entire history of a particle. 
Although a detection with the telescopes reveals the particle aspect at that very moment, it is illegitimate to infer that the past trajectory was that of a straight line \cite{Roussel}. Put differently: complementarity should be applied at the detectors, not at the slits \cite{Srikanth}. 
%does not allow to infer that the past has been that of a travelling point. 
More precise: we can only understand Wheeler's experiment correctly by treating the wave function as fundamental (particle-and-wave view) and not by holding on to strict complementarity (particle-or-wave view)  \cite{Jaques07}.

\begin{figure}
\begin{center}
\includegraphics[width=80mm]{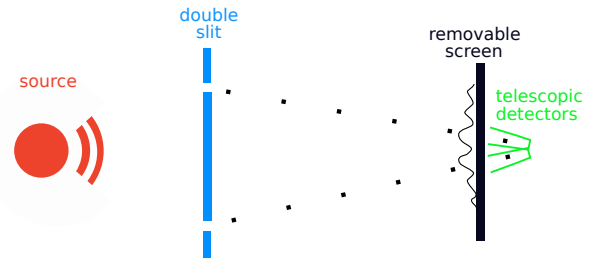}
%wheeler removable screen
\caption{Wheeler's though experiment. %In the double slit experiment, make the detecting screen retractable and place two telescopes behind it, each aimed at one slit. 
When the screen is there, an interference pattern is seen (wave aspect/wavy line). When the screen is pulled away, only one of the two detectors receives a signal (particle aspect/dotted trajectories). %The crux of the thought experiment lies in asking what happens if one pulls away the screen after photon passed the slits. You expect the telescopes to see the particle aspect. But this means your decision (pulling away the screen) influences the behaviour of the photon in the past - namely the moment on which it passed the slits.
}
\label{fig:Wheeler}
\end{center}
\end{figure}
Let us try to understand the experiment at hand with these remarks in mind. The wave function of the particle on the right of the slits is given by the sum of two spherical waves:
\be
\psi(\vec{r}) = \frac{1}{|\vec{r}-\vec{r}_1|} e^{i k |\vec{r}-\vec{r}_1|} + \frac{1}{|\vec{r} - \vec{r}_2|} e^{i k |\vec{r}-\vec{r}_2|}
\label{eq:psie}
\ee
Here $k$ is the wavenumber of the particle; $\vec{r}_1$ and $\vec{r}_2$ are the locations of the two slits.
In the far-away limit, the prefactors $1/|\vec{r}-\vec{r}_1|$ and $1/|\vec{r}-\vec{r}_2|$ are identical up to subleading corrections, and %. Also, the difference in distances can be written as
\be
|\vec{r} - \vec{r}_2|-|\vec{r}-\vec{r}_1| \sim d \sin \theta %\sim d \frac{h}{L}
\ee
where $d$ is the distance between the slits and $\theta$ the angle associated to the position $\vec{r}$ on the screen. 
With these approximations, we get
\be
\psi(\vec{r}) %\sim \frac{1}{r_1} ( e^{i k r_1} + e^{i k r_2}) 
\sim  \frac{1}{|\vec{r}-\vec{r}_1|}  e^{i k |\vec{r}-\vec{r}_1|} (1 + e^{i k d \sin \theta})
\ee
%Which reproduces the well-known result that 
So constructive interference occurs at points where $k d \sin \theta = 2 n \pi$. But what do we see when we pull the screen away? %When we are using 
The (sharply aimed) telescopes actually measure the \textit{momentum} of the incoming particle,  
%The size of the momentum is irrelevant, but the direction precisely tells through which slit the particle came. So the two telescopes are nothing but 
%Or better: the detectors measure the \textit{direction} of the momentum, 
each with a very narrow detection range. For positions $\vec{r}$ near to the location $\vec{r}_T$ of the telescopes, we can approximate (\ref{eq:psie}) as a sum of two planar waves: %(pure momentum states) 
%around $\vec{r}$, % (location of the detectors):
%. This is easy: in the far-away limit, the two spherical waves are (locally) very close to planar ones, so we can approximate
\be
\psi(\vec{r})\sim  \frac{1}{|\vec{r}-\vec{r}_1|} e^{i \vec{k}_1 (\vec{r}-\vec{r}_1)} + \frac{1}{|\vec{r}-\vec{r}_2|} e^{i \vec{k}_2 (\vec{r}-\vec{r}_2)}
\ee
%Here $\vec{r}_1$ and $\vec{r}_2$ are the locations of the two slits, and $\vec{r}$ is the location of the detectors. %are the vectors pointing from each slit to the location $\vec{r}$ of the telescopes behind the screen. 
where the vectors $\vec{k}_1$ and $\vec{k}_2$ are wavevectors of size $k$, directed along $\vec{r}_T-\vec{r}_1$ and $\vec{r}_T-\vec{r}_2$  respectively. Since the telescopes %above formula says exactly what we want. When we aim a telescope at the first slit, it 
are set to detect momenta within a small interval around $\vec{k}_1$ and $\vec{k}_2$ respectively, the above expression implies that their detection probabilities are proportional to %the amplitudes of these momentum components. These read 
$1/|\vec{r}-\vec{r}_1|^2$ and $1/|\vec{r}-\vec{r}_2|^2$. Since these are identical in the far-away limit, %) identical up to subleading corrections. 
the two telescopes have equal chances to detect the particle, just as we expected.

This simple understanding of Wheeler's experiment may lead one to doubt whether it truly is a case of delayed choice. This is the reason why we have dealt with the (more clear-cut) quantum eraser in the main text.

\newpage

\section*{Appendix B: Everettian view}
So far, the discussions in this note have been entirely independent from the specific wave function collapse mechanism. We have only relied on the Born rule, independent of its underlying origin. In this appendix we assume (as an illustration) the Everettian view on the (apparent) wave function collapse. This gives a concrete implementation of the elements we have met so far, and might provide a more intuitive understanding of the observations made above.
 
To introduce the Everettian interpretation, one usually starts with the notion of premeasurement. Say we have a particle which can be in two states: $|\uparrow\rangle$ and $|\downarrow\rangle$. Also, we have a measuring device (initially in state $|I\rangle$) which can detect the state of the particle. We denote the total system (device+particle) \textit{after} the measurement by $|\textrm{UP}\rangle$ or $|\textrm{DOWN}\rangle$, depending on which outcome is shown in the display of the device. Since the device should do an honest job, we want the time evolution to work as follows:
\begin{eqnarray}
 |\uparrow\rangle \otimes |I\rangle &\rightarrow& |\textrm{UP}\rangle \\
 |\downarrow\rangle \otimes |I\rangle &\rightarrow& |\textrm{DOWN}\rangle 
\end{eqnarray}
By linearity this necessarily means that 
\be
(\alpha |\uparrow\rangle + \beta |\downarrow\rangle)\otimes |I\rangle \rightarrow (\alpha |\textrm{UP}\rangle + \beta|\textrm{DOWN}\rangle)
\ee
So if the particle starts out in a superposition, the measuring device has to end up in a superposition too. This clearly conflicts with reality: we never see an entire measuring device in a superposition. The explanation of Everett is that the macroscopic difference between the device states $|\textrm{UP}\rangle$ and $|\textrm{DOWN}\rangle$ implies that they can not interfere with each other anymore in the future. Effectively, they are `disconnected' parts of the total wave function. This means that they are like separated worlds \cite{Everett}. To an observer, it looks like the state has collapsed on one particular outcome, although in fact there are two worlds: one in which the observer sees the `up' outcome, and one in which he sees the `down' outcome. (To see the role of the observer more explicitly: re-read the above two formulae again, but now interpret $|I\rangle$ as the initial state of the device+observer and $|\textrm{UP}\rangle$ and $|\textrm{DOWN}\rangle$ as the final state of the particle+device+observer.)

Let us now see how we can understand the main conclusion of this paper in this framework. For concreteness, consider the EPR setting again.%, namely two separated observers and a particle in the state $(|\uparrow\rangle_A|\downarrow\rangle_B+|\downarrow\rangle_A|\uparrow\rangle_B)/\sqrt{2}$. 
We denote the state of observer Alice and her measuring apparatus as $|\textrm{Alice}\rangle$ and similarly for $|\textrm{Bob}\rangle$. So initially, the total state is
\be
\frac{|\uparrow\rangle_A|\downarrow\rangle_B+|\downarrow\rangle_A|\uparrow\rangle_B}{\sqrt{2}}|\textrm{Alice}\rangle |\textrm{Bob}\rangle
\label{eq:split}
\ee
If Alice measures, and at some other (later or earlier) moment Bob, the above state becomes:
\be
%\frac{|\uparrow\rangle_A|\downarrow\rangle_B}{\sqrt{2}} |\textrm{Alice: up}\rangle |\textrm{Bob: down}\rangle+\frac{|\downarrow\rangle_A|\uparrow\rangle_B}{\sqrt{2}}|\textrm{Alice: down}\rangle |\textrm{Bob: up}\rangle
\frac{1}{\sqrt{2}} |\textrm{Alice: up}\rangle |\textrm{Bob: down}\rangle+\frac{1}{\sqrt{2}}|\textrm{Alice: down}\rangle |\textrm{Bob: up}\rangle
\label{eq:aftersplit}
\ee
Here $|\textrm{Alice: up}\rangle$ denotes her measuring device indicating up, and Alice observing this result. (And similarly for the other combinations.)
So this again suggests a world splitting event like before. But we can also view this another way. Rewrite (\ref{eq:split}) as
\be
\frac{|\uparrow\rangle_A|\downarrow\rangle_B}{\sqrt{2}}|\textrm{Alice}\rangle |\textrm{Bob}\rangle+\frac{|\downarrow\rangle_A|\uparrow\rangle_B}{\sqrt{2}}|\textrm{Alice}\rangle |\textrm{Bob}\rangle
\ee
In this form, it is clear that we can view the initial state as \textit{already} being a superposition of two different worlds. In each of these worlds lives a couple Alice and Bob, but they are ignorant of the world they are in. Only when they measure, they discover which of the two worlds they are in. This should make very clear that the relative moment at which they do this discovery does not matter at all. Alice and Bob can choose (at any time they want) to discover the world they are in and from this, they can infer what the other will measure (the opposite) \textit{whenever} he or she chooses to do so. (This once again shows that information is physical \cite{Preskill}, but that it does not `travel' as is confusingly suggested in \cite{QI}.)
This view makes three things completely clear. First, it reconfirms that the act of measuring just has implications in the form of conditional probabilities. Second, it makes clear that the relative ordering of spatially separated measurement is completely irrelevant. Third: it shows explicitly that the wave function `collapse' is not a global dynamical event. The splitting of the worlds is not something which happens along a spatial slice, it has been there `from the beginning'. The only thing which changes by measuring is a decrease of ignorance of Alice an Bob, which clearly is a very local event. Actually, the absence of a global \textit{dynamical} collapse event is intimately related to linearity of quantum mechanics. Indeed, separate worlds can never `feel' each other because of linearity.

One last remark: the time dependence of observers and devices is not trivial, since they need not be energy eigenstates. Strictly speaking, the above formulae are to be read as Heisenberg notation (no time dependence in states) and the `transition' (\ref{eq:split})$\rightarrow$(\ref{eq:aftersplit}) just as a change of basis, appropriate to the late-time configuration. This is a bit sloppy, but the main goal was not rigor, just providing another (more intuitive) view on the conclusions of this paper.


\newpage
\begin{thebibliography}{99}
 
%\bibitem{lamport94}
 % Leslie Lamport,
  %\emph{\LaTeX: A Document Preparation System}.
 % Addison Wesley, Massachusetts,
 % 2nd Edition,
 % 1994.
\bibitem{EPR} A. Einstein, N. Rosen, B. Podolsky, ``Can quantum-mechanical description of physical reality be considered complete?" \textit{Phys. Rev.} \textbf{47}, 777 (1935) % http://scholar.google.com/scholar?hl=nl&client=safari&rls=en&q=author:%22Einstein%22+intitle:%22Can+quantum-mechanical+description+of+physical+reality+...%22+&um=1&ie=UTF-8&oi=scholarr         en     ref in http://www.drchinese.com/David/Bell_Compact.pdf
\bibitem{Bohm} D. Bohm, ``Quantum Theory'', Prentice-Hall, Englewood Cliffs, 1951.
\bibitem{Bell}  J. S. Bell, ``On the problem of hidden variables in quantum mechanics'', Rev. Mod. Phys. \textbf{38}, 447 (1966)
\bibitem{Photonwave function}  I. Bialynicki-Birula, ``On the wave function of the photon". \textit{Acta Physica Polonica} A 86: 97Ð116 (1994)\\
G.G. Lapaire, J.E. Sipe, ``Photon wave functions and quantum interference experiments",  [quant-ph/0607008] 
\bibitem{Eraser} Y. Kim, R. Yu, S.P. Kulik, Y.H. Shih, M.O. Scully, ``Delayed choice quantum eraser", \textit{Phys. Rev. Lett.}, Vol. 84, No. 1, Jan. 4, 2000. [quant-ph/9903047] 
\bibitem{Wheeler} J.A. Wheeler, ``Mathematical Foundations of Quantum Theory", Academic Press (1978)%ok, is wheelers boek. editor niet vermeld.
\bibitem{ExpWheeler}  V. Jacques, E. Wu, F. Grosshans, F. Treussart, P. Grangier, A. Aspect, J.-F. Roch, ``Experimental realization of WheelerÕs delayed-choice Gedanken Experiment", \textit{Science} \textbf{315}, 966Ð968. [quant-ph/0610241v1] %wheeler uitgevoerd. ook in science
%\bibitem{Wheeler} J. A. Wheeler, pp.182-213 in \textit{Quantum Theory and Measurement}, J. A. Wheeler and W. H. Zurek edit. (Princeton University Press, 1984).  later dan t andere boek...
\bibitem{BTI} P.J. Werbos, L. Dolmatova, ``The Backwards-Time Interpretation of Quantum Mechanics - Revisited With Experiment", [quant-ph/0008036] %heeft precies alleen n e prin
%\bibitem{CondProb} -\rangle VIND M NIET MEER
\bibitem{ShortTime} A. Cho, ``After a Short Delay, Quantum Mechanics Becomes Even Weirder", 
[http://news.sciencemag.org/sciencenow/2007/ 02/16-04.html] %lijkt geen wetenschapper te zijn
\bibitem{QI} R. Garisto ``What is the speed of quantum information?", [quant-ph/0212078v1]\\%speed of quantum information
D. Salart, A. Baas, C. Branciard, N. Gisin, and H. Zbinden, ``Testing spooky action at a distance", \textit{Nature} \textbf{454}, 861-864, [quant-ph/0808.3316v1] \\%zowel op arxiv als in science
J. Kofler, R. Ursin, C. Brukner, A. Zeilinger, ``Comment on: Testing the speed of Ôspooky action at a distance", [quant-ph/0810.4452]\\
%M. Sato, ``Proposal of Signaling by Interference Control of Delayed-Choice Experimental Setup'', [quant-ph/0409059]%Sato en superliminal signaling.
%(The validity of their experimental outcome is contested in the latter)\\
\bibitem{Coleman} S. R. Coleman, ``Quantum Mechanics in Your Face'', lecture at the New England sectional meeting of the American Physical Society, Apr. 9, 1994, %[URL:http://
[www.physics.harvard.edu/about/video.html]
\bibitem{Peres} A. Peres, ``Quantum theory, concepts and methods'', Kluwer Academic Publishers (1993)
\bibitem{Cont} T.J. Herzog, P.G. Kwiat, H. Weinfurter, and A. Zeilinger, Phys. Rev. Lett. \textbf{75} 3034 (1995)
\bibitem{Cont2} G. Scarcelli, Y. Zhou, Y. Shih, ``Random Delayed-Choice Quantum Eraser via Two-Photon Imaging'', [quant-ph/0512207] 
(and references therein)
\bibitem{Peres} A. Peres, ``Delayed choice for entanglement swapping'',  [quant-ph/9904042]
\bibitem{Sakurai} J. J. Sakurai, ``Modern Quantum Mechanics'', Addison-Wesley, Massachusetts, 1994
\bibitem{Boyle} C.F. Boyle, R.L. Schafir, ``A delayed-choice thought-experiment with later-time entanglement'', [quant-ph/0107098]%beetje zelfde als peres. 
\bibitem{Srikanth} R. Srikanth, ``A quantum field theoretic description of the delayed choice experiment'', [quant-ph/0106154]
%\bibitem{Sellers} W.Sellers, ``A Logical Interpretation of a Delayed-Choice, Quantum Eraser Experiment'', [quant-ph/0303036]%-> validiteit van quant mech beschrijving hier benadrukt. gebruikt wel stopwatches... oh-oh
\bibitem{Roussel} P. Roussel, I. Stefan, ``Is the interpretation of the delayed-choice experiment misleading?'', [quant-ph/0706.2596]
\bibitem{Jaques07} V. Jacques, E. Wu, F. Grosshans, F. Treussart, A. Aspect, Ph. Grangier and J.-F. Roch, ``WheelerÕs delayed-choice thought experiment: Experimental realization and theoretical analysis'', [quant-ph/0710.2597]
%\bibitem{Refs}
\bibitem{Preskill} J. Preskill, ``Quantum Computation'', Chapter 3 [URL: http://www.theory.caltech.edu/people/\\preskill/ph219/\#lecture]
\bibitem{Everett} Hugh Everett, ``Relative state formulation of quantum mechanics'', Rev. Mod. Phys. \textbf{29}: 454Ð462, 1957
%\bibitem{Quantum optics}  q opt
\end{thebibliography}
\end{document}